# Dynamical Mechanism of Two-Dimensional Plasmon Launching by Swift Electrons

Xiao Lin[1,2], Xihang Shi[2], Fei Gao[2], Ido Kaminer[3], Zhaoju Yang[2], Zhen Gao[2], Hrvoje Buljan[4], John D. Joannopoulos[3], Marin Soljačić[3], Hongsheng Chen[1,3*], Baile Zhang[2,5*]

[1]*State Key Laboratory of Modern Optical Instrumentation, Zhejiang University, Hangzhou 310027, China.*
[2]*Division of Physics and Applied Physics, School of Physical and Mathematical Sciences, Nanyang Technological University, Singapore 637371, Singapore.*
[3]*Research Laboratory of Electronics, Massachusetts Institute of Technology, Cambridge, MA, USA.*
[4]*Department of Physics, University of Zagreb, Bijenička c. 32, 10000 Zagreb, Croatia.*
[5]*Centre for Disruptive Photonic Technologies, Nanyang Technological University, Singapore 637371, Singapore.*

**Abstract:** Launching of surface plasmons by swift electrons has long been utilized in electron-energy-loss spectroscopy (EELS) to investigate plasmonic properties of ultrathin, or two-dimensional (2D), electron systems. However, its spatio-temporal process has never been revealed. This is because the impact of an electron will generate not only plasmons, but also photons, whose emission cannot be achieved at a single space-time point, as fundamentally determined from the uncertainty principle. Here, we propose that such a space-time limitation also applies to surface plasmon generation in EELS experiment. On the platform of graphene, we demonstrate within the framework of classical electrodynamics that the launching of 2D plasmons by an electron's impact is delayed after a hydrodynamic splashing-like process, which occurs during the plasmonic "formation time" when the electron traverses the "formation zone." Considering this newly revealed process, we show that previous estimates on the yields of graphene plasmons in EELS have been overestimated.

Since the first successful confirmation of surface plasmons [1-2] on the platform of thin metal films, swift electrons have long been used to investigate plasmonic properties of ultrathin, or two-dimensional (2D), electron systems, including graphene plasmons recently [3-8]. Electron-energy-loss-spectroscopy (EELS), which uses swift electrons as a probe, has been an indispensable tool in studying 2D plasmons [1,3]. On the other hand, despite the long history of studies in 2D plasmons [1-2], the basic mechanism of how a swift electron launches 2D plasmons has never been clearly discussed.

This is because the impact of an electron will generate not only surface plasmons, but also photons (the so called "transition radiation" [9,10]), whose emission cannot be achieved at a single space-time point, but demands a finite space-time region. Historically, the concepts of "formation time" and "formation zone" (i.e., "it takes a relatively long time and therefore a long distance for an energetic electron to create a photon," as retailed in Ref. [11]) was firstly presented by Ter-Mikaelian in Landau's seminar in 1952 [12]. Landau at first strongly opposed these concepts [12], but soon realized their correctness and significance, and further developed them by laying his name in the Landau-Pomeranchuk-Migdal effect, which was experimentally confirmed forty years later. Indeed, the existence of "formation time" and "formation zone" describes our ignorance on the exact moment and location at which a photon is generated. In fact, in the early days of quantum uncertainty principle, Bohr already commented on the impossibility of describing electron-photon interaction "without considering a finite space-time region" [13]. Later, it is Ginzburg and colleagues who included the "formation time" and "formation zone" with mathematical form into the classical framework of transition radiation (established by Ginzburg and Frank in 1945), while still admitting that "comparatively little is known" [9,10].

Although the above space-time discussions in the context of photon emission have lasted for decades, similar discussions have never been performed on the generation of surface plasmons in EELS experiments. This explains the reason why the spatio-temporal process of 2D plasmon launching by a swift electron still remains elusive after a long research history.

In this Letter, we introduce the concepts of "formation time" and "formation zone" into surface plasmon generation in EELS. On the platform of a graphene monolayer, we show within the framework of classical electrodynamics the dynamical process of 2D plasmon launching by a swift electron impacting on the graphene monolayer (see Supplemental Material for a movie showing this process [14]). We link this dynamical process during the "formation time" of graphene plasmons to the deep-water hydrodynamic splashing phenomenon, in which a picosecond jet-like rise of excessive charge concentration is formed on graphene as an analogue of the "Rayleigh jet" (also called "Worthington jet") in hydrodynamic splashing [15-16]. In this newly revealed physical process, a significant part of electromagnetic energy has already been dissipated before graphene plasmons are generated. In view of this consideration, we show that previous estimates of graphene plasmon yields [3-7] from the EELS spectra are overestimated. Note that although we adopt graphene as the platform, our analysis is general in any 2D electron system. In addition, while the formal similarity of dispersion between hydrodynamic water waves and 2D electron systems [17-21] (including graphene plasmons [22-24]) has been investigated with some hydrodynamic-wave-like phenomena predicted [25-28], the analogue to the splashing phenomenon has never been discussed.

The model of calculation is schematically shown in Fig.1. We consider a swift electron with charge $q$, moving with a velocity $\bar{v} = \hat{z}v = \hat{z}\beta c$, where $c$ is the light speed in vacuum. Since the energy loss of the electron coupled to plasmons and transition radiation is much lower than the electron's kinetic energy, the electron's velocity is treated as constant [10]. The space-time dependence of the current density due to the electron is classically described as $\bar{J}^q(\bar{r},t) = \hat{z}J_z^q(\bar{r},t) = \hat{z}qv\delta(\bar{r} - \bar{v}t)$, in which $t = 0$ when the electron goes through the origin. A graphene layer is located at the interface between medium 1 ($z < 0$) with permittivity $\varepsilon_{1r}\varepsilon_0$ and medium 2 ($z > 0$) with permittivity $\varepsilon_{2r}\varepsilon_0$, where $\varepsilon_0$ is the permittivity of vacuum. To simplify analysis, we set medium 1 and medium 2 to be vacuum. As a typical 2D electron system, graphene can be characterized by two macroscopic variables [22]: the deviation of the electron density from its average value $\delta n(\bar{r},t)$, and the associated current density $\delta\bar{J}(\bar{r},t)$ which depends on graphene's surface conductivity $\sigma_s$. In this work, we calculate graphene's surface conductivity based on Kubo formula [29-30] (see details in [14]) and set the chemical potential in graphene $\mu_c = 0.4$ eV (close to the experimentally achievable value [31]). Because of the rotational symmetry, all fields must be transvers magnetic, and thus can be characterized by an electric field component $E_z$ in the $\hat{z}$ direction. By decomposing all the quantities into Fourier components in time and in the coordinates $\bar{r}_\perp = \hat{x}x + \hat{y}y$ with corresponding wavevectors $\bar{\kappa}_\perp = \hat{x}\kappa_x + \hat{y}\kappa_y$, we have [14]

$$J_z^q(\bar{r},t) = \int j_{\bar{\kappa}_\perp,\omega}^q(z) e^{i(\bar{\kappa}_\perp \cdot \bar{r}_\perp - \omega t)} d\bar{\kappa}_\perp d\omega \tag{1}$$

$$E_z(\bar{r},t) = \int E_{\bar{\kappa}_\perp,\omega}(z) e^{i(\bar{\kappa}_\perp \cdot \bar{r}_\perp - \omega t)} d\bar{\kappa}_\perp d\omega \tag{2}$$

By writing the fields due to the electron as $E_{\bar{\kappa}_\perp,\omega}^q$, and the radiation fields in regions of $z<0$ and $z>0$ as $E_{\bar{\kappa}_\perp,\omega}^1$ and $E_{\bar{\kappa}_\perp,\omega}^2$, we can get the total fields in regions of $z<0$ and $z>0$ as $E_{\bar{\kappa}_\perp,\omega}^q + E_{\bar{\kappa}_\perp,\omega}^1$ and $E_{\bar{\kappa}_\perp,\omega}^q + E_{\bar{\kappa}_\perp,\omega}^2$, respectively. Derivation from matching boundary conditions at the graphene plane shows phase relations that $E_{\bar{\kappa}_\perp,\omega}^q \sim \exp(i\omega z/v)$, and $E_{\bar{\kappa}_\perp,\omega}^{1,2} \sim \exp(\mp i\omega z/c\sqrt{1-\kappa_\perp^2 c^2/\omega^2})$, where "$\mp$" in the exponent corresponds to "1,2" in the superscript of $E_{\bar{\kappa}_\perp,\omega}^{1,2}$, and $\kappa_\perp^2 = \kappa_x^2 + \kappa_y^2$ [14].

Ginzburg estimated the length of formation zone (denoted as "formation length" $L_f$) for photon emission in transition radiation based on the following considerations [10]. Inside the formation zone, the total energy of the fields $E_{\bar{\kappa}_\perp,\omega}^q + E_{\bar{\kappa}_\perp,\omega}^{1,2}$ is proportional to $(E_{\bar{\kappa}_\perp,\omega}^q + E_{\bar{\kappa}_\perp,\omega}^{1,2})^2$. The length $L_f$ describes a boundary at which the total energy becomes practically equal to the sum of the energy of the electron field, i.e. $(E_{\bar{\kappa}_\perp,\omega}^q)^2$, and the energy of the radiation field, i.e. $(E_{\bar{\kappa}_\perp,\omega}^{1,2})^2$, meaning that the electron field and the radiation field are separated from each other. In other words, the interference term $E_{\bar{\kappa}_\perp,\omega}^q \cdot E_{\bar{\kappa}_\perp,\omega}^{1,2}$ must play a trivial role. Ginzburg thus set the phase difference of $2\pi$ between $E_{\bar{\kappa}_\perp,\omega}^q$ and $E_{\bar{\kappa}_\perp,\omega}^{1,2}$ to determine the length $L_f$. In other words, $\frac{\omega L_f}{v} \pm \frac{\omega L_f}{c\sqrt{1-\frac{\kappa_\perp^2 c^2}{\omega^2}}} = 2\pi$, which gives

$$L_{f1} = \frac{2\pi}{|\frac{\omega}{v} + \omega/c\sqrt{1-\frac{\kappa_\perp^2 c^2}{\omega^2}}|}, \quad L_{f2} = \frac{2\pi}{|\frac{\omega}{v} - \omega/c\sqrt{1-\frac{\kappa_\perp^2 c^2}{\omega^2}}|} \tag{3}$$

where $L_{f1}$ and $L_{f2}$ correspond to the length of the formation zones in regions of $z<0$ and $z>0$, respectively (see Fig.1). Apparently, the formation length depends on not only the frequency of emitted photons and the swift electron's velocity, but also the emission angle of photons. As a numerical example, with the electron's velocity $v=0.8c$, in order to emit a photon at frequency 10 THz in the backward $-\hat{z}$ direction ($\kappa_\perp^2=0$), it takes the electron about $\frac{L_{f1}}{v} = 0.06$ ps to accomplish it. For photon emission in the forward $+\hat{z}$ direction ($\kappa_\perp^2=0$) at the same frequency, it takes about $\frac{L_{f2}}{v} = 0.5$ ps. These time intervals of 0.06 ps and 0.5 ps are considered as "formation times" for the backward and forward photon emission at 10 THz.

Ginzburg's estimation only applies to photon emission with the condition $\kappa_\perp^2 < \omega^2/c^2$ in order to validate the square root in Eq. (3). Yet it is well-known that surface plasmons have $\kappa_\perp^2 > \omega^2/c^2$. Compared to the electron-photon interaction as analyzed by Ginzburg, in the interaction between the electron and surface plasmons, only the electron field has phase variation along the z axis, while the surface plasmons have zero phase variation. Therefore, along the line of Ginzburg's thought, the phase difference of $2\pi$ between $E^q_{\overline{\kappa}_\perp,\omega}$ and $E^{1,2}_{\overline{\kappa}_\perp,\omega}$ should only come from $E^q_{\overline{\kappa}_\perp,\omega}$, i.e. $\frac{\omega L_f}{v} = 2\pi$. This gives

$$L_{f1} = L_{f2} = \frac{2\pi v}{\omega} \qquad (4)$$

Numerically, it means that it will take the electron about $(L_{f1} + L_{f2})/v = 0.2$ ps to launch a 2D plasmon at 10 THz. This 0.2 ps is considered as the "formation time" for the 2D plasmon at 10 THz.

It is interesting to see what happens during the formation time of 2D plasmons when the electron goes through the formation zone, as schematically illustrated in Fig. 1. The movie in the Supplemental Material shows the dynamical process of photon emission and 2D plasmon launching by a swift electron moving with $v = 0.8c$ (or kinematic energy 340 keV), which can be conveniently realized in modern electron microscope systems [1,32]. (This relatively large velocity also shows that our theoretical analysis is not limited to low speeds of the electron.) The bandwidth of calculation is from 0 to 20 THz that can be justified in the Supplemental Fig.3 [14]. In Fig. 2 we extract some figures to show the temporal evolution of photon emission. When the incident electron moves close to the graphene interface [Fig.2(a)], its evanescent field touches graphene, expelling the conduction electrons in graphene from the vicinity of the swift electron's trajectory. Due to electromagnetic induction, the induced surface current in turn blocks the penetration of the evanescent electron field by accumulating fields on the upper side of graphene [Fig.2(a-b)]. Immediately after the electron crosses the graphene [Fig.2(b-c)], the fields previously accumulated on the upper side of graphene have lost direct contact with the electron, and thus are being "shaken off" into the upper space. Meanwhile, the insufficient fields on the lower side of graphene need to recover their strength in presence of the electron, and thus "shakes off" radiation into the lower space [Fig. 2(c-d)]. It can be seen that, at the central frequency 10 THz of the calculation bandwidth, the formation time of 0.06 ps for the backward radiation is consistent with Fig. 2(a-b), and that of 0.5 ps for the forward radiation is consistent with Fig. 2(c-d).

What is more interesting is the charge motion on graphene. We plot in Fig.3 the dynamics of the deviation of electron density from its average value $\delta n(\bar{r}, t)$ on graphene during the penetration of the swift electron. Similarities to the hydrodynamic splashing scenario caused by a little droplet falling on a 2D liquid surface are evident, as we will elaborate in the following.

First, in hydrodynamics [15-16], when a small droplet impacts a calm deep-water surface, a crater is formed first and water splashes to the side. As the water rushes back to fill the crater, a central jet column, called Rayleigh jet due to its instability [15] or Worthington jet because A. M. Worthington performed extensive observations back to 1908 [15-16], will rise above the initial water surface along the collapse axis. When it comes to charge motion on graphene, a "crater" of the density of conduction electron in graphene is first formed when the incident electron approaches the graphene layer (Fig.3a-c). Immediately following the penetration of the electron through the graphene layer, a rebound of the density of conduction electrons is formed with a central jet-like rise (Fig.3d-e), being analogous to the hallmark of Rayleigh jet or Worthington jet in hydrodynamics. We observe, as in Fig. 3(e), that the jet-like rise reaches its peak at about t=0.15 ps.

Second, after the central jet-like rise falls down, ripples of 2D plasmons propagate outward as concentric circles, as shown in Fig.3(g-h). The excited 2D plasmons, which cover a broad spectrum of frequencies, gradually spread into a sequence of plasmonic ripples where longer-wavelength plasmons stay at the outer periphery and shorter-wavelength plasmons stay closer to the inner boundary (Fig.3h). This is because the dispersion of 2D plasmons (including graphene plasmons) is typically $\omega \propto \sqrt{\kappa_\perp}$ [22-24], which is formally analogous to the dispersion of deep-water-waves; this is reflected in the fact that longer wavelengths go out faster [14], although they are in fact formed later (for their longer "formation times") than the shorter wavelengths.

We plot in Fig. 4(a) the time evolution of total radiated photon energy by numerically monitoring the Poynting power going through an imaginary sphere with a large radius $R$ centered at the origin and then shifting the time axis backwards by $R/c$. The total radiated photon energy eventually saturates and approaches a value of $0.171 \times 10^{-4}$ eV that is analytically obtained by letting $t \to \infty$ [14]. It can be seen that most photon energy has been radiated out before $t = 0.15$ ps, which is the moment when the jet-like rise of charge reaches its maximum [Fig. 3(e)], as most photon energy is "shaken off" by the charge jet. We also plot the time evolution of "energy" of induced fields without considering the interference fields of swift electron. Note that 2D plasmons have not been generated before their formation times. So this "energy" should be treated as a parameter related to the field strength, $(E^{1,2}_{\bar{\kappa}_\perp,\omega})^2$, of induced charges, but not as the real energy. It can be seen that even before most photon energy is radiated out, the field strength of $(E^{1,2}_{\bar{\kappa}_\perp,\omega})^2$ has already started to drop.

Now we calculate the real energy of generated graphene plasmons by checking the electromagnetic energy for each frequency component immediately after the time $t = L_{f2}/v$. For the calculation spectrum of 2D plasmons from 0 to 20 THz, the corresponding formation time $(L_{f1} + L_{f2})/v$ varies from $\infty$ to 0.1 ps, whose duration is comparable with the propagation time (~0.5 ps) for graphene plasmons [14]. The energy loss during the formation time cannot be taken into account by simply resorting to the propagation time or propagation length of 2D plasmons that are widely used to characterize the propagation of graphene plasmons, because this energy loss happens before the generation of 2D plasmons themselves.

Only in the ideal lossless situation can one equal the EELS spectrum to the energy spectrum of generated 2D plasmons (the energy of emitted photons generally occupies only <5% of the energy loss of the swift electron and thus is negligible). We plot in Fig. 4(b) the spectrum of 2D plasmons in the ideal lossless situation by taking $t \to \infty$, which can be considered equivalent to the EELS calculation [14], and that for the realistic lossy graphene after formation time is taken into account by taking $t = L_{f2}/v$. The comparison between the lossless and lossy cases shows that a significant portion of energy (over 25% total energy from 0 to 20 THz) has already been dissipated before 2D plasmons are generated.

In conclusion, we propose that the space-time limitation takes effect in surface plasmon generation, demanding a finite formation time and formation zone. The dynamical mechanism of 2D plasmon launching on the platform of graphene in EELS experiment is demonstrated, where an intermediate process between the impact of the swift electron and the formation of 2D plasmons, which is analogous to the hydrodynamic splashing phenomenon, is revealed. Explicit calculation that takes into account the energy dissipation before 2D plasmons are generated has shown that previous estimates on the yields of graphene plasmons in EELS under the lossless assumption have been overestimated.

**ACKNOWLEDGEMENTS**

The authors thank Prof. Sir John B. Pendry for suggestions on transition radiation, and Dr. Ling Lu for helpful discussions. This work was sponsored by the NNSFC (Grants No. 61322501, No. 61275183), the National Program for Top-Notch Young Professionals, the NCET-12-0489, the FRFCU (2014XZZX003-24), the Innovation Joint Research Center for Cyber-Physical-Society System, Nanyang Technological University for NAP Start-Up Grant, the Singapore Ministry of Education under Grant No. MOE2015-T2-1-070 and MOE2011-T3-1-005, the U. S. Army Research Laboratory and the U. S. Army Research Office through the Institute for Soldier Nanotechnologies (Contract No. W911NF-13-D-0001), the MIT S3TEC EFRC of DOE (Grant No. de-sc0001299), the UKF grant 5/13, and the Chinese Scholarship Council (CSC 201506320075).

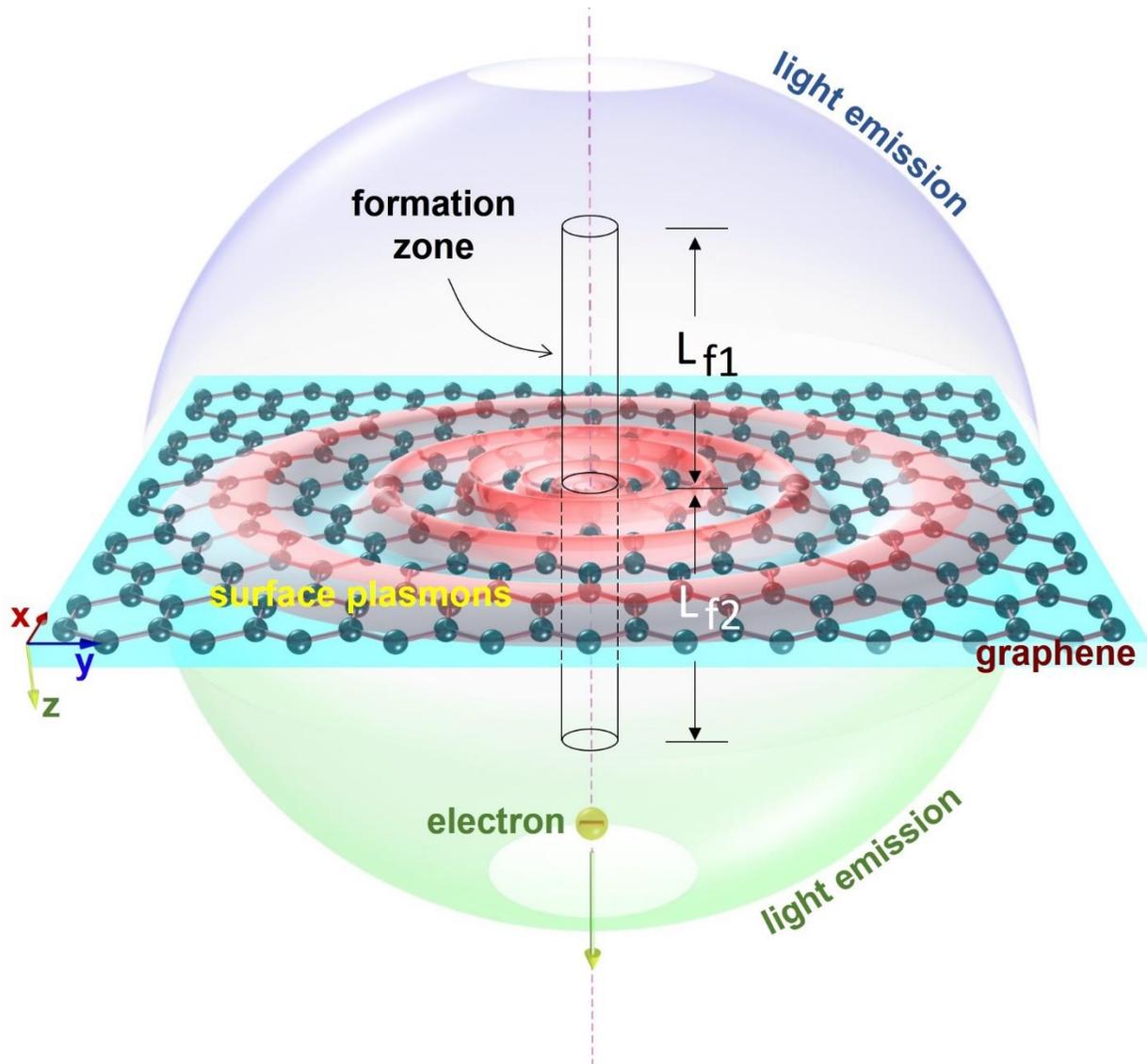

FIG. 1. Schematic of two-dimensional plasmon launching with a swift electron penetrating through a graphene monolayer. $L_{f1}$ and $L_{f2}$ are the length of the formation zone in the region above and below graphene layer, respectively.

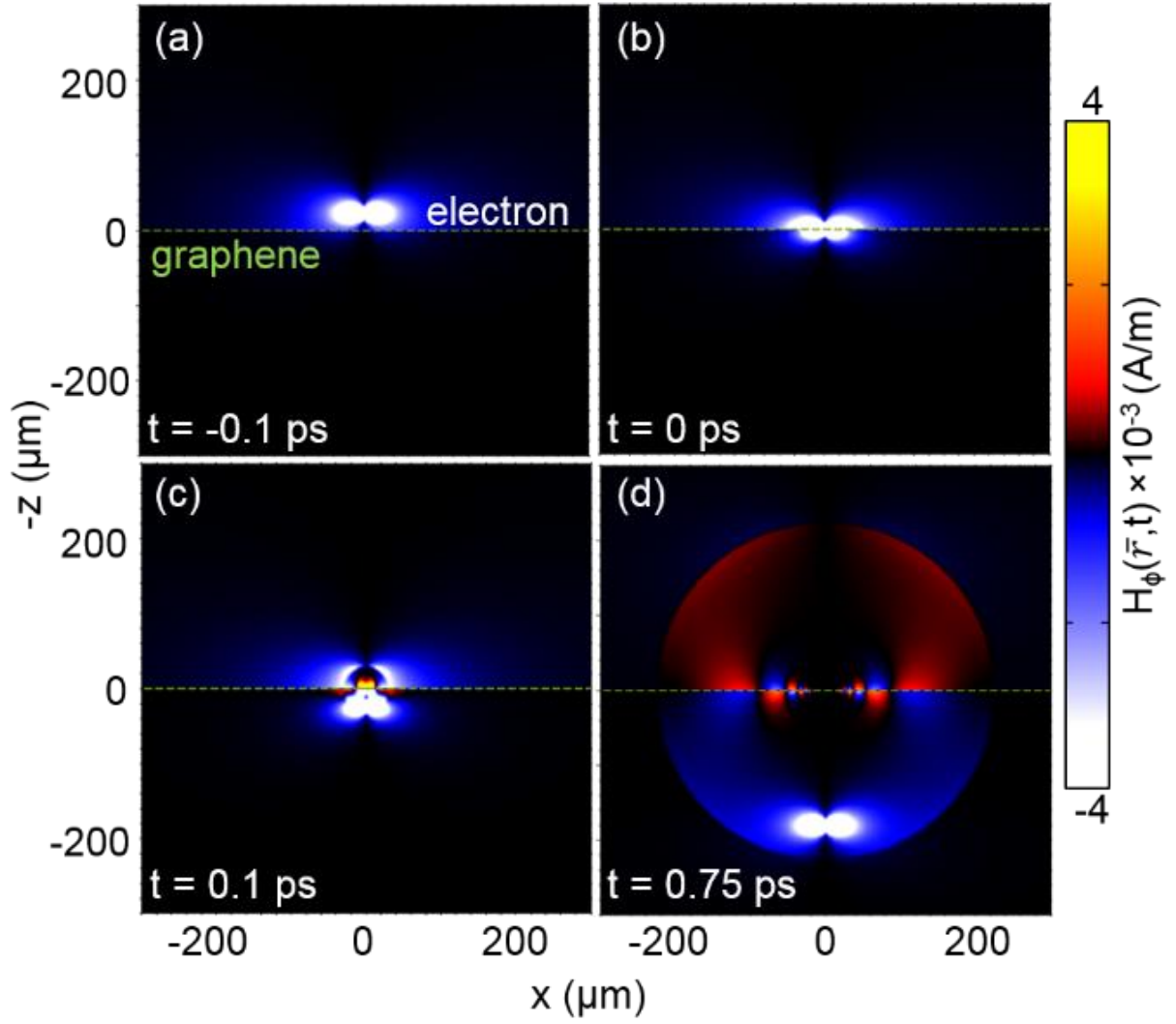

FIG. 2. Time evolution of magnetic field $H_\phi(\vec{r},t)$ when a swift electron perpendicularly penetrates through a graphene monolayer. The green dashed line represents graphene.

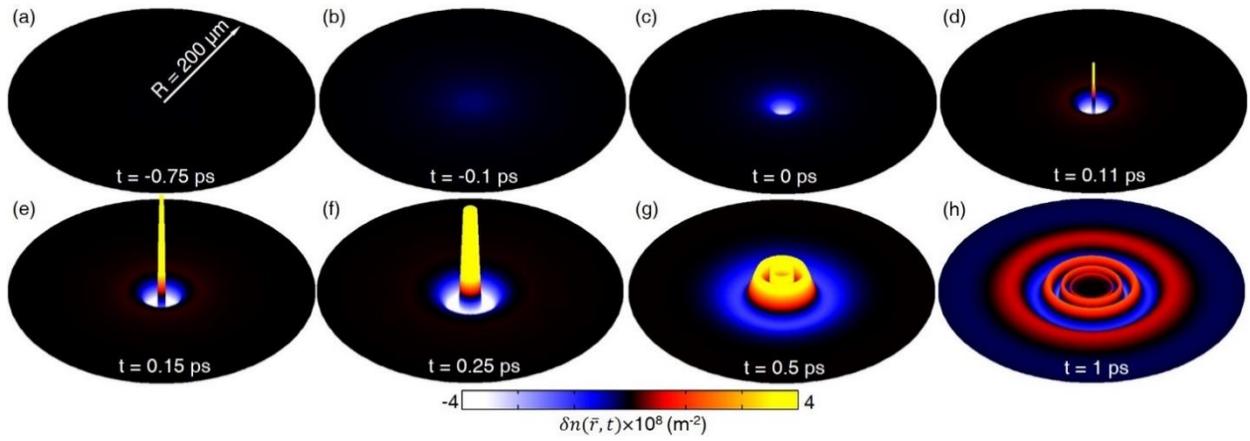

FIG. 3. Time evolution of the deviation of the electron density from its average value on graphene plane $\delta n(\vec{r}, t)$ when a swift electron penetrates through a graphene monolayer.

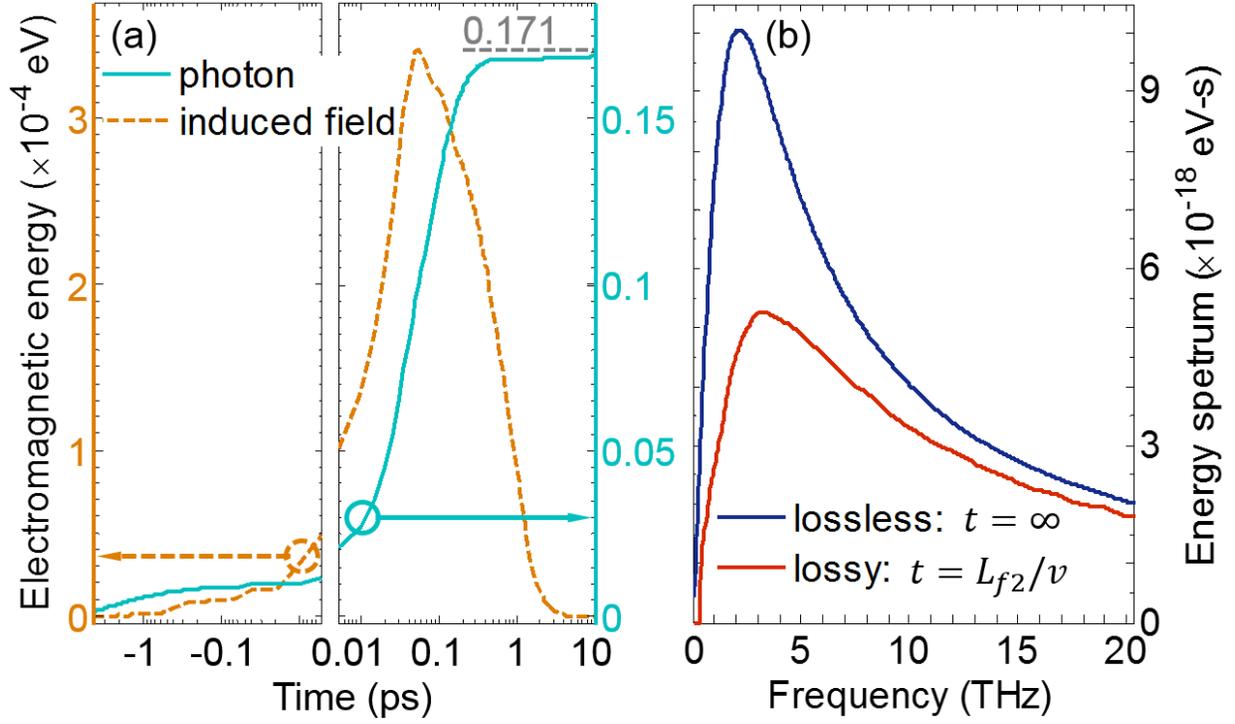

FIG. 4. Energy dissipation during plasmonic formation time. (a) Time evolution of emitted photon energy and the induced field strength $(E^{1,2}_{\bar{\kappa}_\perp,\omega})^2$. (b) Energy spectra of graphene plasmons by taking $t \to \infty$ in the lossless case and by taking $t = L_{f2}/v$ in the lossy case.